# Cosmic Explorer: The U.S. Contribution to Gravitational-Wave Astronomy beyond LIGO

**Principal Author:**
Name: David Reitze
Institution: LIGO Laboratory, California Institute of Technology
Email: dreitze@caltech.edu
Phone: +1–626–395–6274

**Co-authors:** Rana X Adhikari,[a] Stefan Ballmer,[b] Barry Barish,[a] Lisa Barsotti,[c]
GariLynn Billingsley,[a] Duncan A. Brown,[b] Yanbei Chen,[d] Dennis Coyne,[a] Robert Eisenstein,[c]
Matthew Evans,[c] Peter Fritschel,[c] Evan D. Hall,[c] Albert Lazzarini,[a] Geoffrey Lovelace,[e]
Jocelyn Read,[e] B. S. Sathyaprakash,[f] David Shoemaker,[c] Joshua Smith,[e] Calum Torrie,[a]
Salvatore Vitale,[c] Rainer Weiss,[c] Christopher Wipf,[a] Michael Zucker[a,c]

[a]LIGO Laboratory, California Institute of Technology, Pasadena, California 9ll25, USA
[b]Department of Physics, Syracuse University, Syracuse, NY 13244, USA
[c]LIGO Laboratory, Massachusetts Institute of Technology, Cambridge, Massachusetts 02l39, USA
[d]Caltech CaRT, Pasadena, CA 9ll25, USA
[e]California State University Fullerton, Fullerton, CA 9283l, USA
[f]The Pennsylvania State University, University Park, PA 16802, USA

# 1 Executive Summary

Gravitational-wave astronomy is a completely new way to observe the universe. The breakthroughs made by the National Science Foundation's Advanced LIGO, and its partner observatory Advanced Virgo,[1,2] are only the beginning of our exploration of the gravitational-wave sky.[3–7] This white paper describes the research and development that will be needed over the next decade to realize "Cosmic Explorer," the U.S. component of a future third-generation detector network.[8] Cosmic Explorer, together with a network of planned and proposed observatories spanning the gravitational-wave spectrum, including LISA[9,10] and the Einstein Telescope,[11] will be able to determine the nature of the densest matter in the universe; reveal the universe's binary black hole population throughout cosmic time; provide an independent probe of the history of the expanding universe; explore warped spacetime with unprecedented fidelity; and expand our knowledge of how massive stars live, die, and create the matter we see today.

This white paper presents a technology development program that will lead to a two-stage plan for Cosmic Explorer, similar to the path successfully followed by the National Science Foundation's LIGO. The first stage (CE1) scales up Advanced LIGO technologies to create an L-shaped interferometric detector with arms that are closer to the wavelength of the gravitational waves targeted by ground-based detectors. A facility with 40 km long arms is the baseline for achieving Cosmic Explorer's science goals. The second stage (CE2) upgrades the 40 km detector's core optics using cryogenic technologies and new mirror substrates to realize a full order of magnitude sensitivity improvement beyond Advanced LIGO.[12]

With its spectacular sensitivity, Cosmic Explorer will see gravitational-wave sources across the history of the universe. Sources that are barely detectable by Advanced LIGO will be resolved with incredible precision. The explosion in the number of detected sources — up to millions per year — and the fidelity of observations will have wide-ranging impact in physics and astronomy. By peering deep into the gravitational-wave sky, Cosmic Explorer will present a unique opportunity for new and unexpected discoveries. Operating as part of a world-wide network with the Einstein Telescope,[11] or other possible detectors, Cosmic Explorer will be able to precisely localize sources on the sky,[13,14] coupling gravitational-wave astronomy to electromagnetic and particle astronomy.

After a review of Cosmic Explorer's scientific potential (§2) and an overview of its design (§3), we outline the engineering study that must be completed in order to design and construct Cosmic Explorer (§4.1). This includes designing a vacuum system for two 40 km beam tubes and developing the civil engineering program needed to prepare the facility site. We then describe a program of laboratory research and prototyping to evolve existing LIGO-class detector components and concepts to those needed for Cosmic Explorer (§4.2). We discuss a program of international collaboration to coordinate the construction and operation of a unified third-generation network of gravitational-wave detectors (§5). Finally, we summarize the schedule and cost of these activities (§6).

The new technologies needed to realize Cosmic Explorer (including cost-effective long ultrahigh-vacuum systems, civil engineering studies, new optical materials, and cryogenics) will require a substantial investment in research and design, large-scale prototyping, and tabletop research. A three-year "Horizon Study," funded by the National Science Foundation (PHY–1836814), is underway to lay the groundwork for the activities described in this white paper. However, this is only the first step: further investment at a level of $65.7M (cost category: medium scale ground-based) early in the coming decade will be needed to ensure that CE1 can begin observing in the 2030s, with CE2 to be operational in the 2040s.

# 2   Key Science Goals and Objectives

A series of 2020 Decadal Survey White Papers describes the science case for a third-generation gravitational-wave detector network.[3–7] To achieve these science goals, Cosmic Explorer must push the low-frequency sensitivity limit of the detector down by a factor of two, from 10 Hz to 5 Hz, and push the detector sensitivity well beyond the limits of the LIGO facilities[12] (Fig. 1). The leap in sensitivity between second- and third-generation detectors will take the scientific community from first detections to seeing and characterizing every stellar-mass black hole merger in the universe.

The broad and deep discovery aperture of Cosmic Explorer is a consequence of the wide range of sources in the 5–4000 Hz band of the gravitational-wave spectrum, the vast number of objects that it will detect, and the precision with which a third-generation network will map the gravitational-wave sky. In this section, we highlight some of the key science goals that will be possible once third-generation detectors are observing the universe.

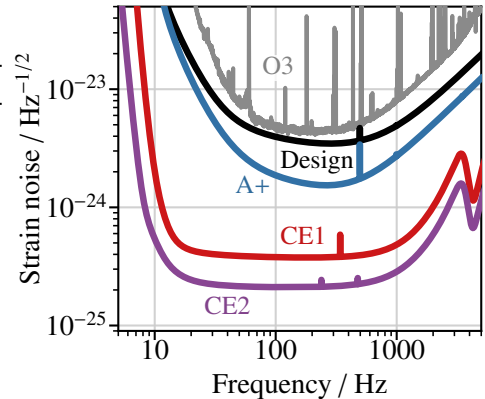

Figure 1: Amplitude spectrum of the detector noise as a function of frequency for the two stages of Cosmic Explorer (CE1, CE2) and the current (O3), design, and upgraded (A+) sensitivities of Advanced LIGO.

### Determining the Nature of the Densest Matter in the Universe

Neutron stars are made of the densest matter in the universe and can have incredibly large magnetic fields. Six decades after discovering neutron stars, we still do not understand how matter behaves at the pressures and densities found in their interiors, or how neutron-star matter can generate magnetic fields a million times stronger than the strongest fields ever created on Earth. Subtle signatures of the star's interior are encoded in the gravitational waves emitted when neutron stars spiral together and merge. Cosmic Explorer will capture these mergers with the precision needed to understand the cold, dense nuclear equation of state that governs neutron-rich matter. When neutron stars merge, they create a hot, dense remnant which can emit both gravitational waves and light in a variety of ways. Observation of the mergers and post-merger remnants of neutron stars can reveal unknown physics in the state of matter at ultra-high densities.

### Multimessenger Observations of Binary Systems

As part of a third-generation network, Cosmic Explorer will provide a unique opportunity to observe thousands of neutron-star mergers in concert with electromagnetic telescopes. A large number of such multimessenger observations is required to understand the dynamics of the collision, the astrophysical processes that form jets, and the formation of heavy elements in the universe. A third-generation network can localize binary neutron stars prior to their merger, allowing electromagnetic facilities to capture the first moments of the collision that are critical to understanding jet physics, kilonovae, and neutrino winds in remnant disks. Some black-hole binaries will also be seen by LISA, allowing us to localize them before they enter the Cosmic Explorer band and look for signs of matter in their circumbinary environment.

### Seeing Black Holes Merge throughout Cosmic Time

Cosmic Explorer can detect merging stellar-mass black holes at redshifts of up to $z \sim 20$. This immense reach will reveal for the first time the complete population of stellar-mass black holes, starting from an epoch when the universe was still assembling its first stars. Cosmic Explorer will detect hundreds of thousands of black-hole mergers each year, measuring their masses and spins. These observations will reveal the black-hole merger rate, the underlying star formation rate, how both have changed throughout cosmic time, and how both are correlated with galaxy evolution.

## Probing the Evolution of the Universe

Observations of gravitational waves from sources at cosmological distances will allow the third-generation network to probe the geometry and expansion of the Universe independent of electromagnetic techniques. We can infer the sources' luminosity distance without the need to calibrate them with standard candles. For some of these sources, electromagnetic counterpart observations will let us measure the source redshift. These observations will let us precisely measure cosmological parameters, such as the Hubble parameter and the dark matter and dark energy densities, giving a completely independent and complementary measurement of the history of the Universe.

## Exploring the Nature of Gravity and Compact Objects

Gravitational waves emanate from spacetime regions of strong gravity and large curvature. The emitted waves encode the nature of the gravitational field, characteristics of the sources, and the physical environment in which they reside. The large number of detected merging black holes will likely include uncommon mergers that are too rare for today's detectors to observe; for example, highly spinning black holes, the inspiral of a neutron star into an intermediate-mass black hole, or a binary black hole with enough surrounding matter to produce an electromagnetic counterpart. Measuring the properties of these rare mergers could revolutionize our understanding of the nature of compact objects, as well as the fundamental nature of gravity. Cosmic Explorer will observe the loudest gravitational-wave sources with an order of magnitude more sensitivity than today's detectors. These extremely high-fidelity observations will put general relativity to the most stringent tests while revealing the nonlinear dynamics of strongly warped spacetime. A third-generation network will have numerous opportunities to discover physics beyond general relativity, for example, in the form of new particles and fields that violate the strong equivalence principle, Lorentz invariance violations, or extra polarizations in addition to the two predicted by general relativity.

## The Life and Death of Massive Stars

The evolution of massive stars, the detailed physics of supernovae, and the origin of pulsar glitches are open problems in astrophysics. Every year, Cosmic Explorer will detect more neutron stars than have been found in radio surveys of our galaxy over the last forty years. Together with a complete census of the universe's stellar-mass black-hole binaries, the properties of this vast population of compact objects will help us understand the evolution of massive stars and the supernova engine that creates the compact-object remnants. The core collapse of a massive star in the Milky Way or Magellanic Clouds will generate gravitational waves that could be directly observed by Cosmic Explorer. Emissions from quakes in pulsars and glitches in magnetars in our galaxy could be detected. Together with electromagnetic-wave and neutrino observations, observations of these systems would revolutionize our understanding of the extreme astrophysics that powers them.

## Sources at the Frontier of Observations

Beyond the known and guaranteed sources of gravitational waves, Cosmic Explorer might see other spectacular sources, the detection of any one of which would be revolutionary. For example, third-generation detectors might detect some forms of dark and exotic matter including axionic and other dark matter fields around black holes or in the cores of neutron stars, the mergers of primordial black holes formed in the early universe, or gravitational-wave emission from cosmic (super)strings.

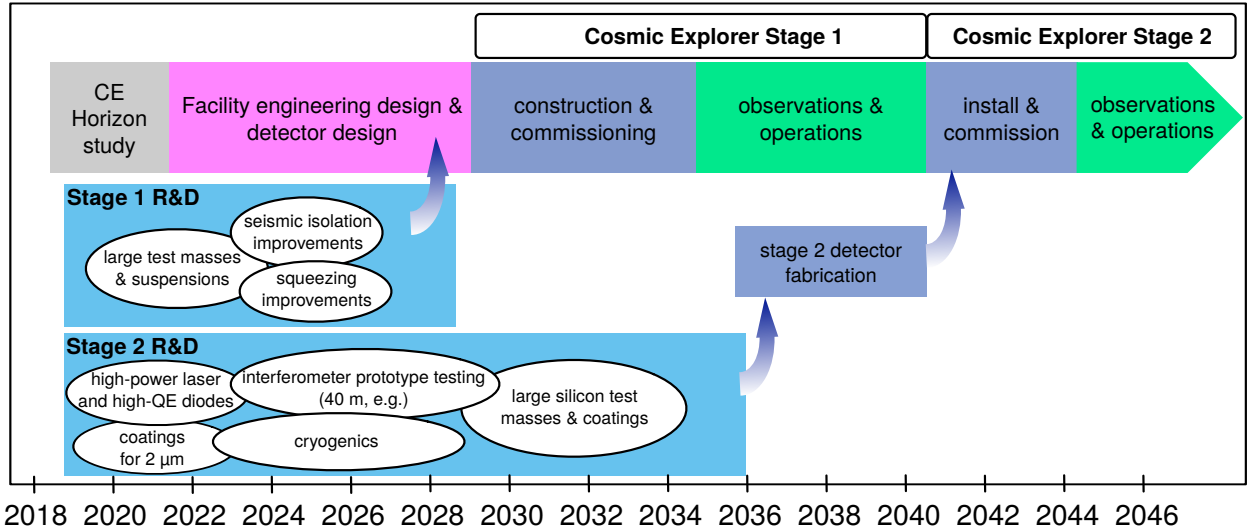

Figure 2: A top-level timeline showing a phased approach to the Cosmic Explorer detector. A detailed timeline for Cosmic Explorer technology development in the 2020s in given in Figure 3.

# 3   Technology Overview

The LIGO and Virgo instruments have opened a new window on the universe, but they are, like Galileo's first telescope, just sensitive enough to observe the brightest sources. Today the Advanced LIGO detectors now see signals roughly weekly; when the recently funded "A+" upgrade comes online in 2024, it will deliver roughly ten detections per week. This may be the most sensitive detector to be installed in the present LIGO infrastructure, and may exhaust the lifetime of the LIGO vacuum systems. The above science goals are only achievable by making observations of these bright sources with significantly higher fidelity and over a wider frequency band, as well as by observing much more distant sources. This requires a new generation of observatories with an order of magnitude greater sensitivity in the audio frequency band than current observatories allow. We envision the U.S. contribution to the global third-generation ground-based gravitational-wave detector network to be Cosmic Explorer, a 40 km L-shaped observatory designed to greatly deepen and clarify humanity's gravitational-wave view of the cosmos.[12]

## 3.1   Design architecture

The Cosmic Explorer facility baseline requirements are two 40 km ultrahigh-vacuum beam tubes, roughly 1 m in diameter, built in an L-shape on the surface of flat and seismically quiet land in the United States. The longer arm length will increase the amplitude of the observed signals with effectively no increase in the noise. Although there are areas of detector technology where improvements would lead to incremental increases in the sensitivity and bandwidth of the instruments, the dominant improvement will come from significantly increasing the arm length.

Cosmic Explorer will be realized in two phases. The initial phase, "Cosmic Explorer Stage 1," is expected to use the technology developed for the "A+" upgrade to Advanced LIGO, scaled up to a 40 km detector with correspondingly better sensitivity. This provides a straightforward approach to significant improvement, as seen in the last rows of Table 1. The second stage, "Cosmic Explorer Stage 2," improves on the Stage 1 sensitivity with a new set of technologies to reduce the quantum and thermal noises of the detector. The main parameters of both stages of Cosmic Explorer, and the comparison with LIGO A+, are given in Table 1, along with a few key astrophysical performance measures. Figure 2 shows the top-level timeline for realizing Cosmic Explorer, with Stage 1 observing in the 2030s and Stage 2 observing in the 2040s. The current NSF-funded "Horizon Study" (PHY–1836814) is further developing the science case for Cosmic Explorer and identifying preliminary considerations for the siting, costing, R&D, and project organization required for Cosmic Explorer to achieve its goals.

|                              | **LIGO A+**           | **CE Stage 1**         | **CE Stage 2**          |
| ---------------------------- | --------------------- | ---------------------- | ----------------------- |
| Arm length                   | 4 km                  | 40 km                  | 40 km                   |
| Test mass                    | 40 kg fused silica    | 320 kg fused silica    | 320 kg silicon          |
| Suspension                   | 0.6 m silica fibers   | 1.2 m silica fibers    | 1.2 m silicon ribbons   |
| Temperature                  | 297 K                 | 297 K                  | 123 K                   |
| Laser wavelength             | 1 µm                  | 1 µm                   | 2 µm                    |
| Circulating power            | 0.8 MW                | 1.4 MW                 | 2 MW                    |
| Squeezed light level         | 6 dB                  | 6 dB                   | 10 dB                   |
| BNS (BBH) horizon redshift[15] | 0.17 (2.6)          | 3.1 (26)               | 12 (37)                 |
| BNS SNR at $z = 0.01$        | 150                   | 1700                   | 3300                    |
| BNS early warning at $z = 0.01$ | 10 minutes        | 40 minutes             | 90 minutes              |

Table 1: Key design parameters and astrophysical performance measures for the LIGO A+ and Cosmic Explorer detectors. The astrophysical performance measures assume a 1.4–1.4$M_\odot$ binary-neutron-star (BNS) system and a 30–30$M_\odot$ binary-black-hole (BBH) system, both optimally oriented. "Early warning" is the time before merger at which the event has accumulated a signal-to-noise ratio (SNR) of 8.

## 3.2   Key performance requirements

Achieving the mid- and high-frequency performance of Cosmic Explorer requires high-quality optical materials, high-power lasers, and careful control of optical losses in the detector. Achieving the low-frequency performance requires eliminating scattered laser light, mechanically isolating the detector test masses from the environment using high-quality multi-stage suspensions and active seismic isolation, and measuring and subtracting local gravity fluctuations induced by the ground and atmosphere.

## 3.3   Technical, site, and infrastructure requirements

The Cosmic Explorer facility must be capable of holding the 40 km beam tubes under ultrahigh vacuum for decades. The facility site and the civil infrastructure must provide as much as possible a stable environment both to preserve the integrity of the beam tubes and to reduce environmental fluctuations near the detector test masses.

# 4   Technology Drivers

Significant technology development is required to realize the Cosmic Explorer facility and detectors. To ensure continuity of gravitational-wave astronomy, it is critical that this development take place in the 2020s, while the current 4 km observatories are still operational.

The construction cost and robustness of the Cosmic Explorer beam tube vacuum system can benefit from recent developments and new ideas in vacuum technology, including novel topologies, materials, treatments, and procedures. Additionally, significant civil engineering is required to prepare the facility site, with one of the main challenges being leveling the 30 m of sagitta attained over 40 km due to the curvature of the Earth. In order to develop these technologies, in the first half of the 2020s we will undertake an engineering study to develop a conceptual design and cost for the vacuum beam tube and the associated civil engineering for an appropriate site; this is described in §4.1.

For the Cosmic Explorer detectors, technology development is required to extend existing LIGO technology to the scale required for Cosmic Explorer Stage 1, principally to develop larger mirrors and to handle the longer arm lengths; most technologies remain unchanged. Parallel development is required to realize the 2 μm cryogenic silicon technologies for Cosmic Explorer Stage 2 (Table 1). Therefore, throughout the 2020s we will undertake a series of laboratory upgrades for large suspensions, cryogenic silicon, and tabletop prototypes as described in §4.2. This R&D effort will identify the most promising approaches for Cosmic Explorer Stage 2, but will not provide a cost estimate for the detector.

## 4.1   An engineering study for Cosmic Explorer

The Cosmic Explorer engineering study will focus on two central and costly aspects of the Cosmic Explorer: (1) the design, fabrication, testing, and maintenance of the 80 km of ultrahigh-vacuum beam tubes, and (2) the civil engineering required to prepare the site to support and lay out the beam tubes.

### 4.1.1   Prerequisite work for the engineering study

Before embarking on the formal engineering study, we will develop a pre-conceptual beam tube vacuum design and identify a plausible reference site for Cosmic Explorer. Together these will serve as a starting point for the engineering study. The pre-conceptual design of the beam tube vacuum system will be based on our experience with the current systems, augmented by recent developments in vacuum technology and by research that we will pursue in the next two years into materials, coatings, pumping strategies and new types of vacuum valves and system geometries. The intent of these studies is to significantly reduce the costs per length of the beam tube vacuum system relative to the initial costs of the l6 km of the present LIGO detectors. The research will include laboratory prototyping of the more promising ideas for cost reduction. The pre-conceptual design will also specify the vacuum pressures, beam tube diameter, and beam tube optical properties to achieve the Cosmic Explorer sensitivity. It will also set operational requirements for pumpdown times, mean time between failures, and allowable motion induced by the environment.

Simultaneously we will identify a plausible reference site in the United States for the 40 km, L-shaped Cosmic Explorer facility. The primary challenge is to lay out a plane on the curved Earth with minimum cost at an accessible location with benign environmental conditions (low probability of earthquakes, floods, *etc.*). The plane will have height differences as large as 30 m compared to the Earth geoid, and smaller differences if the site terrain is bowl-shaped. Critical cost factors will be the differential elevation contours of the site and its geotechnical properties. We will carry out preliminary geotechnical measurements and a preliminary survey at a promising site to establish a concept for the layout of the system — *i.e.*, where the beam tube will be above, below and at the surface. This site will serve as the reference site for the engineering design study. Several alternative sites will be identified for comparison but will not be investigated in as much detail.

### 4.1.2  Description of the engineering study

With the pre-conceptual vacuum design and reference site in hand, the formal engineering design study will proceed in two phases.

**Phase 1** will analyze the pre-conceptual vacuum design and make modifications and iterations as considered useful both from an engineering perspective and to reduce costs. In the beginning of the study the new concepts being proposed — for fabrication, materials, coatings, pumping, and configurations to reduce costs — will be reassessed and the ability to implement them with industry will be determined. This phase of the study will develop a refined conceptual design with enough detail to be able to construct a prototype beam tube vacuum system in the laboratory. The prototype would establish the specialized construction techniques (welding, forming, cleaning), materials, and components (valves and pumps) that would be used in the assembly and demonstrate that the vacuum specifications can be achieved in the required pumping times.

Phase 1 will additionally confirm the properties of the reference site with a full topographic survey and geotechnical investigation. It will then make a revised conceptual design of the beam tube layout and structures at the site. This phase of the engineering design ends with a report on the vacuum performance and the techniques used in making the prototype that will be applied to the construction in the field. An assessment of the reference site and, if needed, a comparison with the alternative sites will be made. It will also provide a first-order costing of the vacuum system, the layout of the structures in the field, and the civil work.

**Phase 2** will produce a full design with complete drawings of the vacuum system and the structures and layout at the site. This phase of the study will also provide a schedule and plan for the logistics to acquire the vacuum components and tube manufacture as well as the civil construction at the site. The second phase will also provide an authoritative cost estimate. More detail and a specific list of tasks for both phases of the engineering study is provided online.[16]

### 4.1.3  Schedule and cost of the engineering study

The preparatory work for the engineering study — namely, development of the pre-conceptual beam tube vacuum design and the identification of a reference site — will be carried out in 2019–2022 and funded by a combination of the current NSF Cosmic Explorer "Horizon Study" grant and a new funding request for lab studies of innovative vacuum system ideas. The vacuum studies will be done in collaboration with vacuum engineers at CERN and Fermilab and other participants of an NSF-sponsored workshop on vacuum technology which held its first meeting in January, 2019.[17] These activities will have been completed before the activities proposed in this white paper begin.

Phase 1 of the engineering study will take place in 2022–2023, with an estimated cost of $15M. Phase 2 of the engineering study will take place in 2024–2025, and will deliver the full conceptual design, a schedule for construction, and the authoritative cost estimate for the vacuum system, structures and civil work. The estimated cost for phase 2 of the study is $18M.

## 4.2  Laboratory research and prototypes

### 4.2.1  Cosmic Explorer prototype test mass chamber

Several LIGO detector technologies will need to be scaled up for Cosmic Explorer. Although the concepts for these detector elements are the same or similar to those for Advanced LIGO, the increased scale for Cosmic Explorer requires significant engineering development and testing. The Cosmic Explorer test masses are scaled up by a factor of two in each dimension, and the test mass suspension is twice as long as in Advanced LIGO. The active seismic isolation system must be re-engineered for this larger payload, and outfitted with new sensors to improve the performance.

The larger test masses also present optical fabrication challenges (polishing and coating) that must be proven on the larger scale. To develop and test these technologies, we will design and build a prototype Cosmic Explorer Test Bed, consisting of a test mass chamber, complete with a seismic isolation system, a full-sized test mass and its suspension, and associated hardware.

New engineering design is also required for the test mass vacuum chamber. In addition to scaling up the size, new features are needed that are motivated by past experience and the future uses of the test bed. These include design elements for scattered light baffling, in-situ particulate removal, improved access for personnel, faster pump-down speed; design elements to support cryogenic suspensions, and design of the interface to the seismic isolation system to minimize vibrations.

Initially this test bed will be used for prototyping Cosmic Explorer Stage 1 — the detector based on room temperature, fused silica test masses and 1 μm lasers. Later, the test bed will prototype Cosmic Explorer Stage 2, with cryogenic silicon masses and 2 μm light. In order to support the preliminary and final designs of Cosmic Explorer, this test bed should be in operation in the 2022–2023 time frame. We estimate the cost of the first phase of the test bed at approximately $20M.

### 4.2.2   Conversion of the LIGO 40 m prototype

To demonstrate the 2 μm cryogenic silicon technology, envisioned for Stage 2 of Cosmic Explorer, we will undertake a modest set of upgrades to the LIGO 40 m prototype interferometer located at Caltech. An initial phase of upgrades is already planned in 2020–2021 to install cryogenic infrastructure, silicon test masses, and a 2 μm laser system. This configuration will establish the basic feasibility of the 2 μm cryogenic silicon technology. A second phase of upgrades will demonstrate the low-noise, high-power capabilities of this technology, including the use of monolithic silicon test mass suspensions, the reduction of coating thermal noises, and the mitigation of scattered light.

Successful demonstration of a prototype 2 μm cryogenic silicon interferometer is an important step for validating the Cosmic Explorer Stage 2 design, and will also set the stage for additional validation by deploying 2 μm cryogenic silicon technology in an existing LIGO facility, if desired; this option is termed the "Voyager" upgrade. The second phase of the 40 m upgrade effort will take place during 2022–2024, with an estimated cost of $3M.

### 4.2.3   Large cryogenic suspension testing

To further develop the detector designs that employ cryogenic test masses, we will outfit the LIGO test mass chamber in the existing LIGO Advanced Systems Test Interferometer facility located at MIT with the largest cryogenic suspension compatible with the current Advanced LIGO infrastructure: a 200 kg silicon test mass, cooled to 123 K. There are three main elements of this project: the test mass suspension, designed for much higher mass than the Advanced LIGO suspension (200 kg vs 40 kg); the test mass cryogenic system; the silicon test mass, polished and coated for 2 μm laser light. We expect that the existing seismic isolation system would accommodate the new payload, perhaps with minor modifications.

The results of cryogenic suspension testing would feed into the second phase of the Cosmic Explorer test bed described earlier. This prototyping would also be directly applicable for deploying such technology in a "Voyager" upgrade of an existing LIGO 4 km facility. The time frame for this R&D effort is expected to be 2022–2023, and we estimate its cost at $4M.

### 4.2.4   Tabletop research

Key proposed elements of both stages of Cosmic Explorer must be researched on tabletop before being scaled up and integrated into a kilometer-scale detector. Of particular importance are the technologies to manufacture adequate test masses: low-loss mirror coatings, low-absorption and low-scatter optics, as well as sensing and control of wavefront distortion induced by high-power laser light. Cryogenic detectors additionally require efficient and low-vibration heat extraction from the test masses, and mitigation of cold layer deposition on the coatings. Research is also required for improved quantum noise reduction techniques and their extension to 2 μm laser wavelength, improved inertial sensors for seismic isolation, and low-noise, high-power lasers. Some of these efforts are already underway in the US and worldwide as part of a broader program to advance gravitational-wave detector technology.[18] For tabletop efforts in the US throughout the 2020s directly related to Cosmic Explorer, we anticipate an investment of $3M.

| Activity | | Cost, M$ | Time frame | Estimated by |
|---|---|---|---|---|
| Engineering study phase 1 | (§4.1.2) | 15 | 2022–23 | ⎱ LIGO Laboratory and |
| Engineering study phase 2 | (§4.1.2) | 18 | 2024–25 | ⎰ vacuum consultants |
| Prototype CE chamber | (§4.2.1) | 20 | 2022–23 | LIGO Laboratory |
| Caltech 40 m upgrade | (§4.2.2) | 3 | 2022–24 | LIGO Laboratory |
| MIT LASTI upgrade | (§4.2.3) | 4 | 2023–24 | LIGO Laboratory |
| Tabletop research | (§4.2.4) | 3 | 2022–28 | CE Horizon Study Team |
| CE project planning | (§5.1) | 0.7 | 2022–24 | LIGO Laboratory |
| Global governance | (§5.2) | 2 | 2022–24 | LIGO Laboratory |
| Total | | 65.7 | | |

Table 2: Summary of activities, schedules, and cost estimates. Estimates performed in June 2019.

# 5 Organization, Partnerships, and Current Status

## 5.1 Project planning for Cosmic Explorer

To provide a complete picture of the path to an operating Cosmic Explorer observatory, in parallel with the engineering study we will provide a refined timeline (a resource-loaded schedule) and identification of critical paths showing the major deliverables and milestones (and possible difficulties) that were identified in the study. In addition, we will examine the observing capabilities and plans for LIGO and other second-generation detectors, and the R&D readiness. Phasing of the construction project will be addressed. Possible management plans will be evaluated, exploring the advantages of coordination between global gravitational-wave detectors. Estimates of operating costs for the Observatory will also be developed using LIGO experience as a guide and scaling where appropriate. The costs of exploiting the science, and scoping computational resources and community support, will be estimated. Approximately $700k will be needed for this planning.

## 5.2 Cosmic Explorer and the Global Gravitational-Wave Network

We anticipate that Cosmic Explorer will operate with other third-generation detectors, such as Einstein Telescope, as part of a global network. For the network to operate effectively, it is critical that there be a unified, coordinated worldwide effort to ensure that the science goals can be realized and that investments in each observatory are leveraged by the development of the network. The science requires coordinated development and observing programs, optimally with three third-generation detectors distributed to localize sources on the sky, with the data analyzed as a network to maximize the science return. The ground-based gravitational-wave community has recognized the imperative to form a globally coherent effort, and has made some progress toward that goal, but we will significantly further the worldwide coordination in parallel with the engineering study.

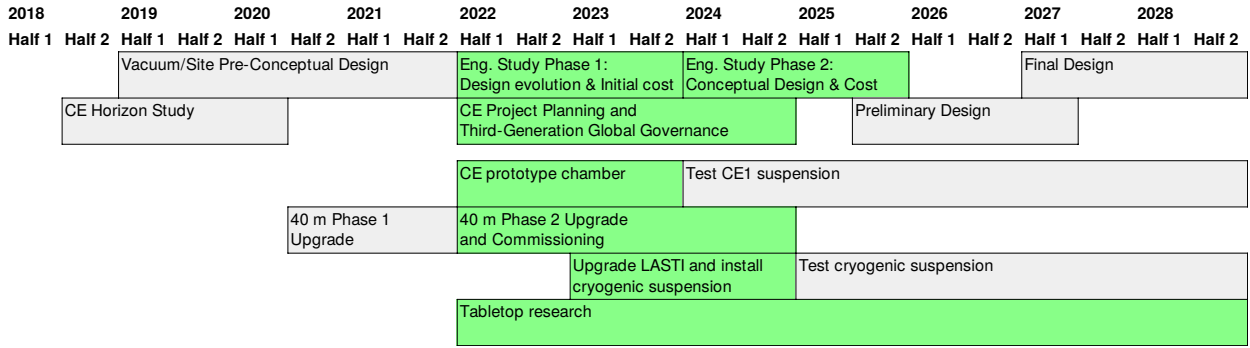

Figure 3: Schedule for Cosmic Explorer technology development in the 2020s. Activities in green are within the scope of this white paper and contribute to the cost estimate (Table 2). The broader timeline for Cosmic Explorer, extending into the 2040s, is given in Figure 2.

Towards this end, a series of NSF-supported meetings[19–22] began in 2015 to plan for the future of ground-based gravitational-wave astronomy. The Gravitational-Wave International Committee (GWIC)[23] chartered a subcommittee to study detector astrophysical and instrumental science.[24] Einstein Telescope has formed a Consortium,[25] and core US participants in the Cosmic Explorer effort are participating in the "Horizon Study" funded by the NSF. We will form an "umbrella organization" for these current endeavors. This organization is starting to take form now, but needs to evolve from a forum for discussion of a coordination of projects and sharing of effort (the current state) to a single worldwide laboratory for gravitational-wave observation. We will develop and realize an operating global governance through interactions with the international community and funding agencies and work to put it in place.

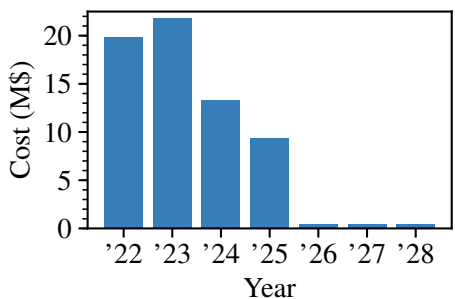

Figure 4: Cost per year for the activities in Table 2 and Figure 3.

To achieve this goal, we see a need for support for community meetings, technical workshops, participation in conferences, extended working visits, organization of documentation and facilitation/organization of design information, and advocacy to and collaboration with the greater scientific community. This effort will serve to continue to cultivate the organization, and create personal relationships ("socialization") between members of different efforts. The end goal in the 2020s is a governance organization that is facilitating the design and implementation of Cosmic Explorer and other third-generation detectors, and that can grow to running the worldwide third-generation observatory network. We estimate $2M over three years is needed for this community development program.

# 6   Schedule and Cost Estimates

A summary of the technological development activities, along with costs and timeframe, is given in Table 2; the total cost for the activities is $65.7M. We intend primarily to solicit federal funding, and possibly also private support. A schedule of these activities, along with other Cosmic Explorer activities in the 2020s, is shown in Figure 3 and the cost breakdown per year is shown in Figure 4.